\documentclass[
aps,%
12pt,%
final,%
notitlepage,%
oneside,%
onecolumn,%
nobibnotes,%
nofootinbib,%
superscriptaddress,%
noshowpacs,%
centertags]
{revtex4}

\usepackage{amsmath,amssymb}
\begin{document}

\title{A superintegrable discrete harmonic oscillator\\
based on bivariate Charlier polynomials}

\author{\firstname{Vincent} X.~\surname{Genest}}
\email{vxgenest@mit.edu}
\affiliation{%
Department of Mathematics, Massachusetts Institute of Technology, 77 Massachusetts Ave, Cambridge, MA 02135 USA 
}%
\author{\firstname{Hiroshi}~\surname{Miki}}
\email{hmiki@mail.doshisha.ac.jp}
\affiliation{%
Department of Electronics, Faculty of Science and Engineering, Doshisha University, Kyotanabe City, Kyoto 610 0394 Japan
}%

\author{\firstname{Luc}~\surname{Vinet}}
\email{vinet@crm.umontreal.ca}
\affiliation{%
Centre de Recherches Math\'{e}matiques, Universit\'{e} de Montr\'{e}al, P. O. Box 6128, Centre-ville Station, Montr\'{e}al, QC, H3C 3J7, Canada
}%
\affiliation{%
Department of Mathematics, Shanghai Jiao Tong University, Shanghai 200240 China
}%

\author{\firstname{Guofu}~\surname{Yu}}
\email{gfyu@sjtu.edu.cn}
\affiliation{%
Department of Mathematics, Shanghai Jiao Tong University, Shanghai 200240 China
}

\begin{abstract}
A simple discrete model of the two dimensional isotropic harmonic oscillator is presented. It is superintegrable with $\mathfrak{su}(2)$ as its symmetry algebra. It is constructed with the help of the algebraic properties of the bivariate Charlier polynomials. This adds to the other discrete superintegrable models of the oscillator based on Krawtchouk and Meixner  orthogonal polynomials in several variables.
\end{abstract}

\maketitle

\section{Introduction}
For various applications, particularly in optical image processing and in signal analysis, it is useful to have discrete models of the harmonic oscillator that preserve most of the properties of the system in the continuum.
Several such discrete models of the one-dimensional quantum oscillator have been developed and the reader might consult ref. \cite{1} for a review.
Of particular relevance to the present paper is the model based on the one-variable Charlier polynomials that is constructed in \cite{2}, see also \cite{3}.
It is desirable to work out models in two dimensions as discrete planar oscillators can be employed for instance to describe wave  guides and pixellated screens.
In general, combining two discrete one-dimensional oscillator models does not lead to a system that has the symmetries of the two-dimensional problem in the continuum \cite{4,5}.
(Interestingly, we shall here have a counterexample to that statement.) 
The isotropic harmonic oscillator is the paradigm example of a maximally superintegrable system.
In $d$ dimensions, such quantum systems with Hamiltonian $H$, possess $2d-1$ algebraically independent constants of motion including $H$.
These quantities that commute with $H$ necessarily form a non-Abelian algebra and in the case of the isotropic harmonic oscillator they generate the Lie algebra $\mathfrak{su}(d)$.    
Superintegrable systems are the hallmark of exactly solvable models, they represent a meaningful laboratory for the study of symmetries and their mathematical description and they prove highly useful in applications.
It is thus of considerable interest to enlarge their catalogue. For a review of (mostly continuous) superintegrable systems, see \cite{6}. 

In the recent past the poor set of known discrete superintegrable models has been significantly increased.
Indeed, two discrete models of the 2-dimensional isotropic harmonic oscillator with $\mathfrak{su}(2)$ symmetry have been obtained \cite{7,8}.
While the two-dimensional situation is described in details in the papers quoted, these models have natural extensions to arbitrary dimension $d$.
Their construction crucially relies on the properties of orthogonal polynomials in $d$ variables that have recently received algebraic interpretations.
In the first case \cite{7}, one called upon the multivariate Krawtchouk polynomials now known \cite{9} to arise as matrix elements of representations of $SO(d+1)$ on oscillator states.
In the second case \cite{8}, use was made of the Meixner polynomials in $d$ variables that are similarly interpreted \cite{10} with the help of the pseudo-orthogonal group $SO(d,1)$.
In that vein, some of us have also given \cite{11} an algebraic description of the Charlier polynomials in $d$ variables based on the Euclidean group in $d$ dimensions.
We here exploit the properties of these polynomials obtained in \cite{11} to construct another discrete model of the isotropic harmonic oscillator thereby complementing the two models we have mentioned.

The outline of the paper is as follows.
In section 2, we will review the features of the bivariate Charlier polynomials that stem from their interpretation in terms of $E(2)$ representations.
In section 3, the discrete model of the isotropic oscillator will be defined and shown to be superintegrable with an $\mathfrak{su}(2)$ invariance algebra.
The wavefunctions will be recorded.
In section 4, it will be confirmed that we have a model of the oscillator by showing that this system is found in the continuum limit.
We shall conclude with some remarks.
An Appendix collecting the relevant properties of the orthonormalized Charlier polynomials in one variable is also included.

\section{The two-variable Charlier polynomials}

The bivariate Charlier polynomials have been seen in \cite{11} to arise in matrix elements of representations of the Euclidean group in two dimensions.
Let $\theta,\alpha$ and $\beta$ be the rotation angle and the translations that define a generic Euclidean transformation of the plane with coordinates $x$ and $y$
\begin{subequations}
\begin{align}
&x \quad \to \quad \cos \theta x+\sin \theta y +\frac{\alpha }{\sqrt{2}},\\
&y \quad \to \quad -\sin \theta x +\cos \theta y +\frac{\beta }{\sqrt{2}}.
\end{align}
\end{subequations}
These will occur as parameters of the bivariate Charlier polynomials $C_{n_1,n_2}(x_1,x_2)$ that can be defined by the generating function
\begin{align}\label{generating:charlier}
\begin{split}
&e^{-z_1(\alpha \cos \theta -\beta \sin \theta )}
e^{-z_2(\alpha \sin \theta +\beta \cos \theta )}
\left[ 1+\frac{z_1}{\alpha }
\cos \theta +\frac{z_2}{\alpha }\sin \theta \right]^{x_1}\left[ 1-\frac{z_1}{\beta }\sin \theta +\frac{z_2}{\beta }\cos \theta \right]^{x_2}\\
&=\sum_{n_1,n_2=0}^{\infty } C_{n_1,n_2}(x_1,x_2) \frac{z_1^{n_1}z_2^{n_2}}{\sqrt{n_1!n_2!}}.
\end{split}
\end{align}
It can be seen from \eqref{generating:charlier} that $C_{n_1,n_2}(x_1,x_2)$ are polynomials of total degree $n_1+n_2$ in the variables $x_1$ and $x_2$ with values in $\mathbb{N}$ ($x_1,x_2=0,1,2,\dots $). They satisfy the orthogonality relation
\begin{equation}\label{orthogonality}
\sum_{x_1,x_2=0}^{\infty }w(x_1,x_2)C_{n_1,n_2}(x_1,x_2)C_{n_1',n_2'}(x_1,x_2)=
\delta_{n_1,n_1'}\delta_{n_2,n_2'},
\end{equation}
where $w(x_1,x_2)$ is the product of two independent Poisson distribution with parameters $\alpha ^2 $ and $\beta^2$:
\begin{equation}\label{weight}
w(x_1,x_2)=e^{-(\alpha^2 +\beta^2)}\frac{\alpha ^{2x_1}\beta^{2x_2}}{x_1!x_2!}.
\end{equation} 
The polynomials $C_{n_1,n_2}(x_1,x_2)$ can be seen from their generating function to have the following explicit expression as an Aomoto-Gel'fand hypergeometric series:
\begin{align}
\begin{split}
C_{n_1,n_2}(x_1,x_2)&=\frac{(-1)^{n_1+n_2}}{\sqrt{n_1!n_2!}}(\alpha \cos \theta-\beta \sin \theta )^{n_1}(\alpha \sin \theta +\beta \cos \theta )^{n_2}\\
&\sum_{\rho,\sigma,\mu ,\nu=0}^{\infty } \frac{(-n_1)_{\rho +\mu}(-n_2)_{\sigma +\nu} (-x_1)_{\rho +\sigma }(-x_2)_{\mu+\nu }}{\rho! \sigma! \mu! \nu !}u_{11}^{\rho} u_{12}^{\sigma }u_{21}^{\mu} u_{22}^{\nu},
\end{split}
\end{align}  
where the parameters $u_{ij}$ are given by 
\begin{align}
\begin{split}
u_{11}&=\frac{-\cos \theta }{\alpha^2 \cos \theta -\alpha \beta \sin \theta },\quad 
u_{12}=\frac{-\sin \theta }{\alpha^2 \sin \theta +\alpha \beta \cos \theta },\\
u_{21}&=\frac{-\sin \theta }{\beta^2 \sin \theta -\alpha \beta \cos \theta },\quad 
u_{22}=\frac{-\cos \theta }{\beta^2 \cos \theta +\alpha \beta \sin \theta },
\end{split}
\end{align}
and where $(a)_n$ stands for 
\begin{equation}
(a)_n=a(a+1)\cdots (a+n-1),\quad (a)_0=1.
\end{equation}
The algebraic description of the multivariate Charlier polynomials allows for a very natural characterization of these functions. It leads in particular to an immediate identification of ladder operators that shall prove essential in the construction the superintegrable model. Let $T_{x_i}^{\pm }f(x_i)=f(x_i\pm 1)$ for $i=1,2$ denote the discrete shift operators in the variable $x_1$ and $x_2$. Introduce the following two operators 
\begin{subequations}\label{raising}
\begin{align}
A_+^{(1)}&=\frac{x_1}{\alpha }\cos \theta T_{x_1}^- - \frac{x_2}{\beta }\sin \theta T_{x_2}^- -(\alpha \cos \theta - \beta \sin \theta ),\\
A_+^{(2)}&=\frac{x_1}{\alpha }\sin \theta T_{x_1}^- + \frac{x_2}{\beta }\cos \theta T_{x_2}^- - (\alpha \sin \theta +\beta \cos \theta ).
\end{align}
\end{subequations}
They act as raising operators on the polynomials $C_{n_1,n_2}(x_1,x_2)$:
\begin{subequations}\label{raising_on_charlier}
\begin{align}
&A_+^{(1)}C_{n_1,n_2}(x_1,x_2)=\sqrt{n_1+1}C_{n_1+1,n_2}(x_1,x_2),\\
&A_+^{(2)}C_{n_1,n_2}(x_1,x_2)=\sqrt{n_2+1}C_{n_1,n_2+1}(x_1,x_2).
\end{align}
\end{subequations}
Introduce similarly the operators $A_{-}^{(1)}$ and $A_-^{(2)}$ defined by 
\begin{subequations}\label{lowering}
\begin{align}
A_-^{(1)}&=\alpha \cos \theta T_{x_1}^+ - \beta \sin \theta T_{x_2}^+ -(\alpha \cos \theta - \beta \sin \theta ),\\
A_-^{(2)}&=\alpha \sin \theta T_{x_1}^+ + \beta \cos \theta T_{x_2}^+ - (\alpha \sin \theta +\beta \cos \theta ).
\end{align}
\end{subequations}
They are the lowering operators:
\begin{subequations}\label{lowering_on_charlier}
\begin{align}
&A_-^{(1)}C_{n_1,n_2}(x_1,x_2)=\sqrt{n_1}C_{n_1-1,n_2}(x_1,x_2),\\
&A_-^{(2)}C_{n_1,n_2}(x_1,x_2)=\sqrt{n_2}C_{n_1,n_2-1}(x_1,x_2).
\end{align}
\end{subequations}
The ladder operators \eqref{raising} and \eqref{lowering} can be combined to produce the two commuting difference operators $Y_1 $ and $Y_2$ that are diagonalized by the bivariate Charlier polynomials. These operators are given by
\begin{equation}
Y_i= A_+^{(i)}A_-^{(i)},\quad i=1,2.
\end{equation}
Explicitly they read
\begin{subequations}
\begin{align}
\begin{split}
Y_1&=-x_1 \frac{\beta}{\alpha }\cos \theta \sin \theta T_{x_1}^- T_{x_2}^+ -x_2 \frac{\alpha }{\beta }\cos \theta \sin \theta T_{x_1}^+T_{x_2}^-\\
& - \omega \cos \theta \left[ \frac{x_1}{\alpha }T_{x_1}^- +\alpha T_{x_1}^+\right] + \omega \sin \theta \left[ \frac{x_2}{\beta }T_{x_2}^- +\beta T_{x_2}^+\right] \\
&+(x_1\cos^2 \theta +x_2 \sin^2 \theta +\omega ^2),
\end{split}\\
\begin{split}
Y_2&=x_1 \frac{\beta}{\alpha }\cos \theta \sin \theta T_{x_1}^- T_{x_2}^+ +x_2 \frac{\alpha }{\beta }\cos \theta \sin \theta T_{x_1}^+T_{x_2}^-\\
& - \zeta \sin \theta \left[ \frac{x_1}{\alpha }T_{x_1}^- +\alpha T_{x_1}^+\right] -\zeta \cos \theta \left[ \frac{x_2}{\beta }T_{x_2}^- +\beta T_{x_2}^+\right] \\
&+(x_1\sin^2 \theta +x_2 \cos^2 \theta +\zeta ^2),
\end{split}
\end{align}
\end{subequations}
where 
\begin{equation}
\omega = \alpha \cos \theta -\beta \sin \theta ,\quad \zeta =\alpha \sin \theta +\beta \cos \theta.
\end{equation}
The eigenvalue equations are 
\begin{equation}
Y_i C_{n_1,n_2}(x_1,x_2)=n_i C_{n_1,n_2}(x_1,x_2)\quad i=1,2.
\end{equation}
Note that they fully determine the polynomials $C_{n_1,n_2}(x_1,x_2)$.
Let us also write down the form of $A_+^{(1)}A_-^{(2)}$ and $A_{+}^{(2)}A_-^{(1)}$ as difference operators 
\begin{subequations}
\begin{align}
\begin{split}
A_+^{(1)}A_-^{(2)}&=x_1\frac{\beta }{\alpha}\cos ^2\theta T_{x_1}^-T_{x_2}^+-x_2\frac{\alpha }{\beta}\sin ^2\theta T_{x_1}^+T_{x_2}^-\\
&-\left[ x_1\frac{\zeta }{\alpha }\cos \theta T_{x_1}^-+\alpha \omega \sin \theta T_{x_1}^+\right] +\left[ x_2\frac{\zeta }{\beta }\sin \theta T_{x_2}^--\beta \omega \cos \theta T_{x_2}^+\right]\\
&+((x_1-x_2)\sin\theta \cos \theta +\omega \zeta),
\end{split}\\
\begin{split}
A_+^{(2)}A_-^{(1)}&=-x_1\frac{\beta }{\alpha}\sin ^2\theta T_{x_1}^-T_{x_2}^++x_2\frac{\alpha }{\beta}\cos ^2\theta T_{x_1}^+T_{x_2}^-\\
&-\left[ x_1\frac{\omega }{\alpha }\sin \theta T_{x_1}^-+\alpha \zeta \cos \theta T_{x_1}^+\right] -\left[ x_2\frac{\omega }{\beta }\cos \theta T_{x_2}^--\beta \zeta \sin \theta T_{x_2}^+\right]\\
&+((x_1-x_2)\sin\theta \cos \theta +\omega \zeta),
\end{split}
\end{align}
\end{subequations}
which act on $C_{n_1,n_2}(x_1,x_2)$ as follows:
\begin{subequations}\label{diffrence_operator}
\begin{align}
A_+^{(1)}A_-^{(2)}C_{n_1,n_2}(x_1,x_2)&=\sqrt{(n_1+1)n_2}C_{n_1+1,n_2-1}(x_1,x_2),\\
A_+^{(2)}A_-^{(1)}C_{n_1,n_2}(x_1,x_2)&=\sqrt{n_1(n_2+1)}C_{n_1-1,n_2+1}(x_1,x_2).
\end{align}
\end{subequations}

\section{A discrete and superintegrable Hamiltonian}
We present in this section a model which is superintegrable with a Hamiltonian that is a difference operator in two variables. We take for the Hamiltonian the sum of the operators $Y_1$ and $Y_2$ introduced in the previous section. We hence posit 
\begin{equation}\label{hamiltonian}
H=Y_1+Y_2.
\end{equation}
A simple calculation shows that $H$ is the following operator:
\begin{equation}\label{hamiltonian_form}
H= -x_1 T_{x_1}^- -\alpha^2 T_{x_1}^+ - x_2 T_{x_2}^- - \beta^2 T_{x_2}^+ +x_1+x_2+\alpha ^2 +\beta^2 .
\end{equation}
Let us remark that this is a rather simple expression, that the operators $T_{x_1}^+ T_{{x_2}^-}$ and $T_{x_1}^-T_{x_2}^+$ have cancelled in the combination $Y_1+Y_2$, that the parameter $\theta $ does not appear and that $H$ is simply the sum of the operators \eqref{one-dim_diff} associated to the difference equations of 2 univariate Charlier polynomials with parameter $\alpha^2$ and $\beta^2$.
As for the Krawtchouk \cite{7} and the Meixner \cite{8} models, it is straightforward to show that $H$ is superintegrable by exhibiting constants of motion built from the ladder operators. Let $J_X,J_Y$ and $J_Z$ be defined as follows
\begin{subequations}\label{contants_of_motion}
\begin{align}
J_X&=\frac{1}{2}(A_+^{(1)}A_-^{(2)}+A_+^{(2)}A_-^{(1)}),\\
J_Y&=\frac{1}{2i}(A_+^{(1)}A_-^{(2)}-A_+^{(2)}A_-^{(1)}),\\
J_Z&=\frac{1}{2}(A_+^{(1)}A_-^{(1)}-A_+^{(2)}A_-^{(2)}).
\end{align}
\end{subequations}
It is immediate to verify using \eqref{raising} and \eqref{lowering} that these operators commute with the Hamiltonian \eqref{hamiltonian}:
\begin{equation}
[H,J_X]=[H,J_Y]=[H,J_Z]=0.
\end{equation}
One also has
\begin{equation}
[J_X,J_Y]=iJ_Z,\quad [J_Y,J_Z]=iJ_X,\quad [J_Z,J_X]=iJ_Y.
\end{equation}
This confirms that $H$ is superintegrable with $\mathfrak{su}(2)$ as its symmetry algebra. In the realization \eqref{contants_of_motion}, the $\mathfrak{su}(2)$ Casimir operator is related to the Hamiltonian \eqref{hamiltonian} through
\begin{equation}
J_X^2+J_Y^2+J_Z^2 = \frac{H}{2}\left( \frac{H}{2}+1\right).
\end{equation}
This of course is totally in keeping with the Schwinger realization of $\mathfrak{su}(2)$ in terms of the creation and annihilation operators of 2 independent oscillators.
We have thus constructed in this case, a discrete superintegrable system by the addition of two discrete uni-dimensional models of the oscillator (see \cite{2}). We shall return to this point in our concluding remarks.  

By construction, the wavefunctions of the Hamiltonian \eqref{hamiltonian} are expressed in terms of the two-variable Charlier polynomials $C_{n_1,n_2}(x_1,x_2)$. These wavefunctions $\Phi_{N,n}(x_1,x_2)$ are labelled by the two non-negative integers $N$ and $n$ and read 
\begin{equation}
\Phi_{N,n}(x_1,x_2)=C_{n,N-n}(x_1,x_2),\quad n=0,\dots ,N,~~N=0,1,2,\dots .
\end{equation}
They are eigenfunctions of $H$ and $J_Z$:
\begin{subequations}
\begin{align}
H\Phi _{N,n}(x_1,x_2)&=N\Phi _{N,n}(x_1,x_2),\\
J_Z\Phi_{N,n}(x_1,x_2)&=\left( n-\frac{N}{2}\right) \Phi_{N,n}(x_1,x_2).
\end{align}
\end{subequations}
The energy eigenvalues are given by the non-negative integers $N=0,1,2,\dots $ and have an $(N+1)$-fold degeneracy. For a fixed energy, that is, for a given $N$, the states $\Phi_{N,n}(x_1,x_2)$ support a $(N+1)$-dimensional irreducible representation of $\mathfrak{su}(2)$. With
\begin{equation}
J_{\pm } = J_X\pm iJ_Y,
\end{equation}  
it is seen from \eqref{diffrence_operator} that
\begin{subequations}
\begin{align}
J_+ \Phi_{N,n} (x_1,x_2)&=\sqrt{(n+1)(N-n)}\Phi_{N,n+1}(x_1,x_2),\\
J_- \Phi_{N,n} (x_1,x_2)&=\sqrt{n(N-n+1)}\Phi_{N,n-1}(x_1,x_2).
\end{align}
\end{subequations} 
It thus follows that the set of all two-variable Charlier polynomials $C_{n_1,n_2}(x_1,x_2)$ such that $K=n_1+n_2$ with $K$ a fixed integer, form a $(K+1)$-dimensional irreducible representation space for $\mathfrak{su}(2)$.

In view of \eqref{orthogonality}, the wavefunctions $\Phi_{N,n}(x_1,x_2)$ are not normalized on the grid $(x_1,x_2)\in \mathbb{N}\times \mathbb{N}$ with respect to the standard uniform measure of quantum mechanics. Properly normalized wavefunctions $\Upsilon_{N,n}(x_1,x_2)$ are obtained by taking
\begin{equation}
\Upsilon_{N,n}(x_1,x_2)=\sqrt{w(x_1,x_2)}C_{n,N-n}(x_1,x_2),
\end{equation} 
where $w(x_1,x_2)$ is given by \eqref{weight}. One has then the orthogonality and completeness relations
\begin{subequations}
\begin{align}
&\sum_{x_1,x_2=0}^{\infty }\Upsilon_{N,n}(x_1,x_2)\Upsilon_{N',n'}(x_1,x_2)=\delta_{nn'}\delta_{NN'},\\
&\sum_{N=0}^{\infty }\sum_{n=0}^N \Upsilon_{N,n}(x_1,x_2)\Upsilon_{N,n}(x_1',x_2')=\delta_{x_1x_1'}\delta_{x_2x_2'}.
\end{align}
\end{subequations}
The form of the Hamiltonian or of any other operator that acts on the normalized wavefunctions $\Upsilon_{N,n}(x_1,x_2)$ is obtained by performing the gauge transformation $X \to w^{\frac{1}{2}}X w^{-\frac{1}{2}}$ on the corresponding operator which has been defined in the Charlier or $\Phi _{N,n}(x_1,x_2)$ basis.
For the Hamiltonian $H$, this will show that the resulting operator is the sum of 2 one-dimensional Charlier Hamiltonians as given in \cite{2}.

\section{Limit to the standard oscillator in the continuum}
It will now be shown by taking the continuum limit that the system provided in the last section is a discrete model of the standard harmonic oscillator.
This will be done first, by observing that the discrete wavefunctions tend to the usual oscillator wavefunctions that involve a product of two Hermite polynomials.
Second we shall indicate that the discrete operators also have the proper limits.
We observed that the Hamiltonian $H$ does not depend on $\theta $.
Upon setting $\theta =0$ in the generating formula \eqref{generating:charlier}, we see with the help of the univariate generating formula \eqref{generating_one_charlier} that
\begin{equation}
C_{n_1,n_2}(x_1,x_2)|_{\theta=0} =(-1)^{n_1+n_2}\alpha ^{n_1}\beta^{n_2} \widehat{C}_{n_1} (x_1)\widehat{C}_{n_2}(x_2),
\end{equation}
where $\widehat{C}_n(x)$ are the orthonormalized Charlier polynomials in one-variable defined in \eqref{normal_one_charlier}.
This separation of variables suggests using the well-known limit from the univariate Charlier polynomials to the usual Hermite polynomials $H_n(x)$.
Set 
\begin{equation}\label{parametrization}
x_1=\sqrt{2}\alpha \tilde{x}_1 +\alpha^2 ,\quad x_2 =\sqrt{2}\beta \tilde{x}_2 +\beta^2 ,\quad \theta =0.
\end{equation} 
One has
\begin{equation}\label{generating_limit}
\lim_{\alpha\to \infty }\lim_{\beta\to \infty } e^{-\alpha z_1}\left( 1+\frac{z_1}{\alpha }\right)^{x_1}e^{-\beta z_2}\left( 1+\frac{z_2}{\beta }\right)^{x_2} = e^{-\frac{z_1^2}{2}+\sqrt{2}\tilde{x}_1 z_1}e^{-\frac{z_2^2}{2}+\sqrt{2}\tilde{x}_2 z_2}.
\end{equation} 
Upon comparing with the generating function of the Hermite polynomials
\begin{equation}
e^{-t^2+2xt}=\sum_{n=0}^{\infty }\frac{H_n(x)}{n!}t^n,
\end{equation}
it is immediate to see from \eqref{generating_limit} that
\begin{equation}
\lim_{\alpha\to \infty }\lim_{\beta \to \infty }C_{n_1,n_2}(x_,x_2)|_{\theta =0}=\frac{1}{\sqrt{2^{n_1+n_2}n_1!n_2!}}H_{n_1}(\tilde{x}_1)H_{n_2}(\tilde{x}_2).
\end{equation}
Moreover, with the parametrization \eqref{parametrization}, it is easily found using Stirling's approximation that the Poisson distribution converges to the Gaussian distribution:
\begin{equation}
\lim_{\alpha \to \infty } e^{-\alpha^2 }\frac{\alpha^{2x_1}}{x_1!}=\frac{e^{-\tilde{x}_1^2}}{\sqrt{\pi}},\quad \lim_{\beta \to \infty } e^{-\beta ^2 }\frac{\beta ^{2x_2}}{x_2!}=\frac{e^{-\tilde{x}_2^2}}{\sqrt{\pi}}.
\end{equation}
Hence the wavefunctions
\[
\Upsilon_{N,n}(x_1,x_2)=\sqrt{w(x_1,x_2)}C_{n,N-n}(x_1,x_2)
\]
tend to the oscillator wavefunctions
\begin{equation}
\phi_{N,n}(\tilde{x}_1,\tilde{x}_2)=\frac{1}{ \sqrt{\pi 2^{N}n!(N-n)!}}e^{-\frac{\tilde{x}_1^2+\tilde{x}_2^2}{2}}H_{n}(\tilde{x}_1)H_{N-n}(\tilde{x}_2).
\end{equation}
This justifies calling \eqref{hamiltonian} a discrete 2-dimensional oscillator.
Now let us remark that it is not necessary to take $\theta =0$ in the limiting process.
 If one keeps $\theta $ arbitrary and performs the change of variables \eqref{parametrization}, one obtains in the limit $\alpha \to \infty ,\beta \to \infty $, a product of Hermite polynomials in the rotated coordinates $\hat{x}_1=\cos \theta \tilde{x}_1-\sin \theta \tilde{x}_2$ and $\hat{x}_2=\sin \theta \tilde{x}_1+\cos \theta \tilde{x}_2$. 
 
Let us now examine what happens to the ladder operators under these limits.
After the gauge transformations $\bar{A}_{\pm }^{(i)}=w^{\frac{1}{2}}(x_1,x_2)A_{\pm}^{(i)}w^{-\frac{1}{2}}(x_1,x_2)$, the raising operators become
\begin{subequations}\label{gauged_raising}
\begin{align}
\bar{A}_+^{(1)}&=\cos \theta \sqrt{x_1} T_{x_1}^- -\sin \theta \sqrt{x_2}T_{x_2}^--(\alpha \cos \theta -\beta \sin \theta ),\\
\bar{A}_+^{(2)}&=\sin \theta \sqrt{x_1} T_{x_1}^- +\cos \theta \sqrt{x_2}T_{x_2}^--(\alpha \sin \theta +\beta \cos \theta )
\end{align}
\end{subequations}
and the lowering operators read
\begin{subequations}\label{gauged_lowerin}
\begin{align}
\bar{A}_-^{(1)}&=\cos \theta \sqrt{x_1+1} T_{x_1}^+ -\sin \theta \sqrt{x_2+1}T_{x_2}^+-(\alpha \cos \theta -\beta \sin \theta ),\\
\bar{A}_-^{(2)}&=\sin \theta \sqrt{x_1+1} T_{x_1}^- +\cos \theta \sqrt{x_2+1}T_{x_2}^--(\alpha \sin \theta +\beta \cos \theta ).
\end{align}
\end{subequations}
Performing again the change of variables \eqref{parametrization} and taking the limit $\alpha\to\infty ,\beta \to \infty $, it is seen expectedly that the operators \eqref{gauged_raising} and \eqref{gauged_lowerin} become the rotated creation/annihilation operators
\begin{align*}
&\bar{A}_+^{(1)}\to \cos \theta  a_1^+-\sin \theta a_2^+,\quad \bar{A}_+^{(2)}\to \sin \theta a_1^+ +\cos \theta a_2^+,\\
&\bar{A}_-^{(1)}\to \cos \theta a_1 -\sin \theta a_2 ,\quad \bar{A}_-^{(2)}\to \sin \theta a_1 +\cos \theta a_2,
\end{align*}
where 
\[
a_i=\frac{\tilde{x}_i+\partial_{\tilde{x}_i}}{\sqrt{2}},\quad a_i^+=\frac{\tilde{x}_i-\partial_{\tilde{x}_i}}{\sqrt{2}}
\]
are the usual ladder operators of the harmonic oscillator. This offers another confirmation that a discretization of the 2-dimensional harmonic oscillator that preserves its superintegrablity and symmetries has been provided.

\section{Concluding Remarks}
Using properties of the bivariate Charlier polynomials, we have constructed a discrete model of the 2-dimensional isotropic harmonic oscillator which is superintegrable with $\mathfrak{su}(2)$ as its symmetry algebra.
This complements analogous models of the oscillator which are based on the bivariate Krawtchouk and Meixner polynomials.
Interestingly the present model connected to the Charlier polynomials has a feature that distinguishes it from the previous ones.
Indeed it is observed that while possessing the full $\mathfrak{su}(2)$ symmetry of its continuum counterpart, the model provided here consists ultimately in the addition of two one-dimensional Charlier oscillators.
One might wonder why this naive construction of a 2-dimensional system preserves the continuum symmetries in this case while this does not happen when one-dimensional Krawtchouk or Meixner ocillators are combined.
The reason lies simply in the fact, noted in the Appendix, that the ladder operators of the univariate Charlier polynomials do not affect their parameter while this is not so for the Krawtchouk or Meixner polynomials.
This said it should be mentioned that even though the Hamiltonian \eqref{hamiltonian} does not depend on $\theta $ and can be constructed by adding the Hamiltonians of two one-dimensional Charlier oscillators, its wavefunctions $\Phi_{N,n}(x_1,x_2)$ do depend on $\theta $ are not a mere product of univariate Charlier polynomials.
As a result the symmetry operators \eqref{contants_of_motion} also depend on $\theta $.
Only in the continuum can the rotation of the symmetry of the oscillator be used to gauge away $\theta $.
Note that the 2-dimensional model that we discussed simply extends to arbitrary dimensions.

Anisotropic discrete oscillators can also be constructed. 
The Hamiltonian 
\[
\tilde{H}=k_1 Y_1 +k_2Y_2
\]
will have $k_1n_1+k_2 n_2~(n_1,n_2=0,1,\dots )$ as spectrum and will exhibit degeneracies whenever the frequency ratio $k_2/k_1$ is an integer.
As in the continuum, these models will be superintegrable with a polynomial symmetry algebra.
Explicitly one will have
\begin{align*}
\tilde{H}
&=(k_2-k_1)x_1\frac{\beta }{\alpha } \cos \theta \sin \theta T_{x_1}^-T_{x_2}^++(k_2-k_1)x_2\frac{\alpha }{\beta }\cos \theta \sin \theta T_{x_1}^+T_{x_2}^-\\
&-(k_1\omega \cos \theta +k_2 \zeta \sin \theta )\left[ \frac{x_1}{\alpha }T_{x_1}^-+\alpha T_{x_1}^+\right]+(k_1\omega \sin \theta -k_2\zeta \cos \theta ) \left[ \frac{x_2}{\beta }T_{x_2}^-+\beta T_{x_2}^+\right]\\
&+k_1(x_1\cos ^2 \theta +x_2 \sin ^2 \theta +\omega ^2 )+k_2 (x_1\sin ^2 \theta +x_2 \cos ^2 \theta +\zeta ^2)
\end{align*}
and the presence of the operators $T_{x_1}^+T_{x_2}^-$ and $T_{x_1}^-T_{x_2}^+$ should be noted in distinction to the isotropic situation. One observes that the use of the  multivariate Charlier polynomials leads to a model that is more general than the one that results from the anisotropic combination of two one-dimensional Charlier oscillators.

It would be of interest in the future to obtain symmetry-preserving discretizations of other exactly solvable and superintegrable models along the lines followed here.   
\subsection*{Acknowledgements}
HM has benefited from the hospitality of the Department of Mathematics of the Shanghai Jiao Tong University where LV gratefully holds a Chair visiting professorship. VXG holds a postdoctoral fellowship from the National Sciences and Engineering Research Council (NSERC) of Canada. HM is supported by JSPS KAKENHI Grant Number 15K17561. The work of LV is supported in part by NSERC. The research of GFY is supported by  the National Natural Science Foundation of China (grant no. 11371251).

\appendix
\section{Properties of the univariate Charlier polynomials}
We shall record here for reference the properties of the Charlier polynomials in one variable that are needed. They can all be obtained from \cite{12}.
Let $C_n(x,a)$ denote the standard Charlier polynomials
\begin{equation}
C_n(x,a)=~_2F_0 \left( \left.\begin{matrix}
-n,-x \\ -
\end{matrix} \right| -\frac{1}{a}\right)
\end{equation}
with $~_2F_0$ the usual hypergeometric series. We shall provide formulas for the orthonormalized polynomials 
\begin{equation}\label{normal_one_charlier}
\widehat{C}_n(x)=\frac{(-a )^n}{\sqrt{n!}}C_n(x,a ^2).
\end{equation}
\begin{itemize}
\item Orthogonality relation
\begin{equation}
\sum_{x=0}^{\infty }\frac{a ^{2x}}{x!}\widehat{C}_n(x)\widehat{C}_{n'}(x)=\delta_{nn'}.
\end{equation}
\item Generating function
\begin{equation}\label{generating_one_charlier}
e^{-az}\left( 1+\frac{z}{a}\right)^x =\sum_{n=0}^{\infty }\widehat{C}_n(x)\frac{z^n}{\sqrt{n!}}.
\end{equation}
\item Ladder operators
\begin{subequations}
\begin{align}
&A_- \widehat{C}_n(x)=\sqrt{n}\widehat{C}_{n-1}(x),\\
&A_-=a(T^+-1).
\end{align}
\end{subequations}
\begin{subequations}
\begin{align}
&A_+ \widehat{C}_n(x)=\sqrt{n+1}\widehat{C}_{n+1}(x),\\
&A_+=-a+\frac{x}{a}T^-.
\end{align}
\end{subequations}
Note that the ladder operators do not affect the parameter $a$ included in the definition \eqref{normal_one_charlier} of $\widehat{C}_n(x)$.
\item Difference equation
\begin{subequations}
\begin{align}
&Y \widehat{C}_n(x)=n\widehat{C}_{n+1}(x),\\
&Y=A_+A_-=-a^2 T^+ +(x^2+a^2)-xT^-. \label{one-dim_diff}
\end{align}
\end{subequations}
\end{itemize}

\end{document}